%% file: main.tex
\documentclass[sigconf, nonacm, screen]{acmart}

\usepackage[noend]{algpseudocode}
\usepackage{float}
\usepackage{subcaption}
\usepackage{tabularx}
\usepackage{booktabs}
\usepackage{algorithm2e}
\usepackage{multirow}
\usepackage{booktabs}
\usepackage{tablefootnote}
\usepackage{threeparttable}
\usepackage{placeins}
\usepackage{threeparttable}

\usepackage{amsmath,amssymb}
\usepackage{pifont}
\usepackage{xcolor}
\usepackage{listings}
\input{code}

\definecolor{pastelyellow}{RGB}{255, 255, 230}
\definecolor{lightorange}{RGB}{255, 223, 186}
\definecolor{orange}{RGB}{255, 140, 0}

\AtBeginDocument{%
  \providecommand\BibTeX{{%
    \normalfont B\kern-0.5em{\scshape i\kern-0.25em b}\kern-0.8em\TeX}}}

\acmPrice{15.00}
\acmISBN{978-1-4503-XXXX-X/18/06}

\begin{document}

\title{Text2SQL is Not Enough: Unifying AI and Databases with TAG} 

\author{
    Asim Biswal\textsuperscript{1,*} \quad 
    Liana Patel\textsuperscript{2,*} \quad 
    Siddharth Jha\textsuperscript{1} \quad 
    Amog Kamsetty\textsuperscript{1} \quad 
    Shu Liu\textsuperscript{1} \quad 
    \\ Joseph E. Gonzalez\textsuperscript{1} \quad 
    Carlos Guestrin\textsuperscript{2} \quad 
    Matei Zaharia\textsuperscript{1} \\
    \normalsize \textsuperscript{1}UC Berkeley \quad
    \textsuperscript{2}Stanford University \\
}

\thanks{*Both authors contributed equally to this research.}

\renewcommand{\shortauthors}{}

\newif\ifcomments
\commentstrue
\ifcomments
    \providecommand{\liana}[1]{{\color{blue}{/* liana: #1 */}}}
    \providecommand{\asim}[1]{{\color{magenta}{/* asim: #1 */}}}
    \providecommand{\amog}[1]{{\color{orange}{/* amog: #1 */}}}
    \providecommand{\joey}[1]{{\color{red}{/* joey: #1 */}}}
    \providecommand{\shu}[1]{{\color{purple}{/* shu: #1 */}}}

\else
    \providecommand{\liana}[1]{}
\fi

\newif\ifgap
\gaptrue
\ifgap
    \providecommand{\gap}[1]{{\color{purple}{/* gap: #1 */}}}
    \providecommand{\sid}[1]{{\color{red}{/* Sid: #1 */}}}
\else
    \providecommand{\gap}[1]{}
\fi

\newif\ifrev
\ifrev
    \providecommand{\rev}[1]{{\color{blue}{#1}}}

\else
    \providecommand{\rev}[1]{{\color{black}{#1}}}

\fi



\newcommand{\heading}[1] {{\hfill\break\noindent{\textbf{\emph{#1}}} }}
\newcommand{\comma}[1] {{\textit{,\space\space}{#1} }}

\input{Sections/Abstract}






\maketitle

\newcommand{\operators}{\emph{semantic operators}}

\input{Sections/Introduction}

\input{Sections/TAG-definition}
\input{Sections/Design-space}

\input{Sections/Evaluation}
\input{Sections/RelatedWork}
\input{Sections/Conclusion}
\input{Sections/Acknowledgements}

\bibliographystyle{ACM-Reference-Format}
\bibliography{citations}

\onecolumn
\appendix
\input{Sections/Appendix}

\end{document}

%% file: code.tex
\definecolor{codegreen}{rgb}{0,0.6,0}
\definecolor{codegray}{rgb}{0.5,0.5,0.5}

\definecolor{backcolour}{RGB}{245,248,250}
\definecolor{emph}{RGB}{166,88,53}
\definecolor{nightblue}{RGB}{9,49,105}
\definecolor{keywords}{RGB}{207,33,46}
\definecolor{lightpurple}{RGB}{130,81,223}

\lstdefinestyle{mystyle}{
    backgroundcolor=\color{backcolour},   
    commentstyle=\color{codegreen},
    keywordstyle=\color{keywords},
    stringstyle=\color{nightblue},
    basicstyle=\ttfamily\footnotesize,
    breakatwhitespace=false,         
    breaklines=true,                 
    captionpos=b,                    
    keepspaces=true,                 
    numberstyle=\small\color{codegray},
    numbers=left,                    
    numbersep=5pt,   
    xleftmargin=0.2cm,
    aboveskip=0.2cm,
    belowskip=0.1cm,
    showspaces=false,                
    showstringspaces=false,
    showtabs=false,                  
    tabsize=2,
    frame=shadowbox,
    emph={},
    emphstyle={\color{lightpurple}},
}

%% file: Sections/Abstract.tex
\begin{abstract}
AI systems that serve natural language questions over databases promise to unlock tremendous value. Such systems would allow users to leverage the powerful reasoning and knowledge capabilities of language models (LMs) alongside the scalable computational power of data management systems. These combined capabilities would empower users to ask arbitrary natural language questions over custom data sources.
However, existing methods and benchmarks insufficiently explore this setting. Text2SQL methods focus solely on natural language questions that can be expressed in relational algebra, representing a small subset of the questions real users wish to ask. Likewise, Retrieval-Augmented Generation (RAG) considers the limited subset of queries that can be answered with point lookups to one or a few data records within the database. We  propose Table-Augmented Generation (TAG), a unified and general-purpose paradigm for answering natural language questions over databases. The TAG model represents a wide range of interactions between the LM and database that have been previously unexplored and creates exciting research opportunities for leveraging the world knowledge and reasoning capabilities of LMs over data. We systematically develop benchmarks to study the TAG problem and find that standard methods answer no more than 20\% of queries correctly, confirming the need for further research in this area. We release code for the benchmark at \url{https://github.com/TAG-Research/TAG-Bench}.
\end{abstract}

%% file: Sections/Introduction.tex
\section{Introduction}
\label{sec:intro}
Language models promise to revolutionize data management by letting users ask natural language questions over data, which has led to a great deal of research in Text2SQL and Retrieval-Augmented Generation (RAG) methods.
In our experience, however (including from internal workloads and customers at Databricks), users'  questions often transcend the capabilities of these paradigms, demanding new research investment towards systems that 
combine the logical reasoning abilities of database systems with the natural language reasoning abilities of modern language models (LMs).

In particular, we find that real business users' questions often require sophisticated combinations of domain knowledge, world knowledge, exact computation, and semantic reasoning. Database systems clearly provide a source of \textbf{\emph{domain knowledge}} through the up-to-date data they store, as well as \textbf{\emph{exact computation}} at scale (which LMs are bad at).

LMs offer to extend the existing capabilities of databases in two key ways.
First, LMs possess \textbf{\emph{semantic reasoning}} capabilities over textual data, a core element of many natural language user queries. 
For example, a Databricks customer survey showed users wish to ask questions like \emph{which customer reviews of product X are positive?}, or \emph{why did my sales drop during this period?}. These questions present complex reasoning-based tasks, such as sentiment analysis over free-text fields
or summarization of trends. LMs are well-suited to these tasks, which cannot be modeled by the exact computation or relational primitives in traditional database systems.

Secondly, the LM, using knowledge learned during model training and stored implicitly by the model's weights, can powerfully augment the user's data with \textbf{\emph{world knowledge}} that is not captured explicitly by the database's table schema. As an example, a Databricks internal AI user asked \emph{what are the QoQ trends for the "retail" vertical?} over a table containing attributes for account names, products and revenue. To answer this query the system must understand how the business defines \emph{QoQ}
(e.g., the quarter over quarter trends from the last quarter to the current quarter or this quarter last year to this quarter this year), as well as which companies are considered to be in the \emph{retail vertical}.  This task is well-suited to leverage the knowledge held by a pre-trained or fine-tuned LM.

Systems that efficiently leverage databases and LMs together to serve natural language queries, in their full generality, hold potential to transform the way users understand their data. Unfortunately, these questions cannot be answered today by common methods, such as Text2SQL and RAG. 
While Text2SQL methods~\cite{yaghmazadeh_sqlizer_2017, zelle_learning_1996, zhang_benchmarking_2024, yu_typesql_2018} are suitable for the subset of natural language queries that have direct relational equivalents, they cannot handle the vast array of user queries that require semantic reasoning or world knowledge. For instance, the previous user query asking \emph{which customer reviews are positive} may require logical row-wise LM reasoning over reviews to classify each as positive or negative. Similarly the question which asks \emph{why sales dropped} entails a reasoning question that must aggregate information across many table entries. 

On the other hand, the RAG model is limited to simple relevance-based point lookups to a few data records, followed by a single LM invocation. This model serves only the subset of queries answerable by point lookups and also fails to leverage the richer query execution capabilities of many database systems, which leaves computational tasks (e.g., counting, math and filtering) to a single invocation of the error-prone LM. 
In addition to being error prone and inefficient at computational tasks, LMs have also been shown to perform poorly on long-context prompts limiting their ability to reason about data at scale in the generation phase of RAG.

We instead propose \textbf{\emph{table-augmented generation (TAG)}} as a unified 
paradigm 
for systems that answer natural language questions over databases. 
Specifically, TAG  defines three key steps, as shown in Figure~\ref{fig:example_tag}. First, the query synthesis step \texttt{syn} translates the user's arbitrary natural language request $R$ to an executable database query $Q$. Then, the query execution step \texttt{exec} executes $Q$ on the database system to efficiently compute the relevant data $T$. 
Lastly, the answer generation step \texttt{gen} utilizes $R$ and $T$, where the LM is orchestrated, possibly in iterative or recursive patterns over the data, to generate the final natural language answer $A$. 
The TAG model is simple, but powerful: it is defined by the following three equations, but captures a wide range of previously under-studied interactions between LMs and databases. 
\begin{align}
\text{Query Synthesis:}\quad & \texttt{syn}(R)\rightarrow Q \label{eq:query_generation} \\
\text{Query Execution:} \quad & \texttt{exec}(Q) \rightarrow T \label{eq:query_execution} \\
\text{Answer Generation:} \quad & \texttt{gen}(R\comma T)\rightarrow A \label{eq:answer_generation}
\end{align}

Notably, the TAG model unifies prior methods, including both Text2SQL and RAG, which represent special cases of TAG and serve only a limited subset of user questions. 

While several prior works address these special cases of TAG, we provide the first end-to-end TAG benchmark composed of a broad set of realistic queries that require LM reasoning and knowledge capabilities. We demonstrate the significant research challenges posed by these types of questions, as well as the promise of efficient TAG implementations. Our evaluation analyzes the vanilla Text2SQL and RAG baselines as well as two stronger baselines, Text2SQL with LM generation and retrieval with LM-based re-ranking. Across a variety of query types, we find each baseline method consistently fails to achieve high accuracy, never surpassing $20\%$ exact match on the benchmark. On the other hand, we implement hand-written TAG pipelines on top of the recent LOTUS runtime \cite{patel_lotus_2024} and find they achieve up to $20 - 65\%$ higher accuracy compared to the baselines. This significant performance gap demonstrates the promise of building efficient TAG systems.

%% file: Sections/TAG-definition.tex
\section{The TAG Model}
\label{sec:tag_def}
We now describe the TAG model, which takes a natural language request $R$ and returns a natural language answer $A$ grounded in the data source. We outline three main steps that TAG systems implement: query synthesis, query execution, and answer generation. We define TAG tractably as a single iteration of these steps, but one can consider extending TAG in a multi-hop fashion.

\begin{figure}[!t]
  \centering
  \includegraphics[width=\linewidth]{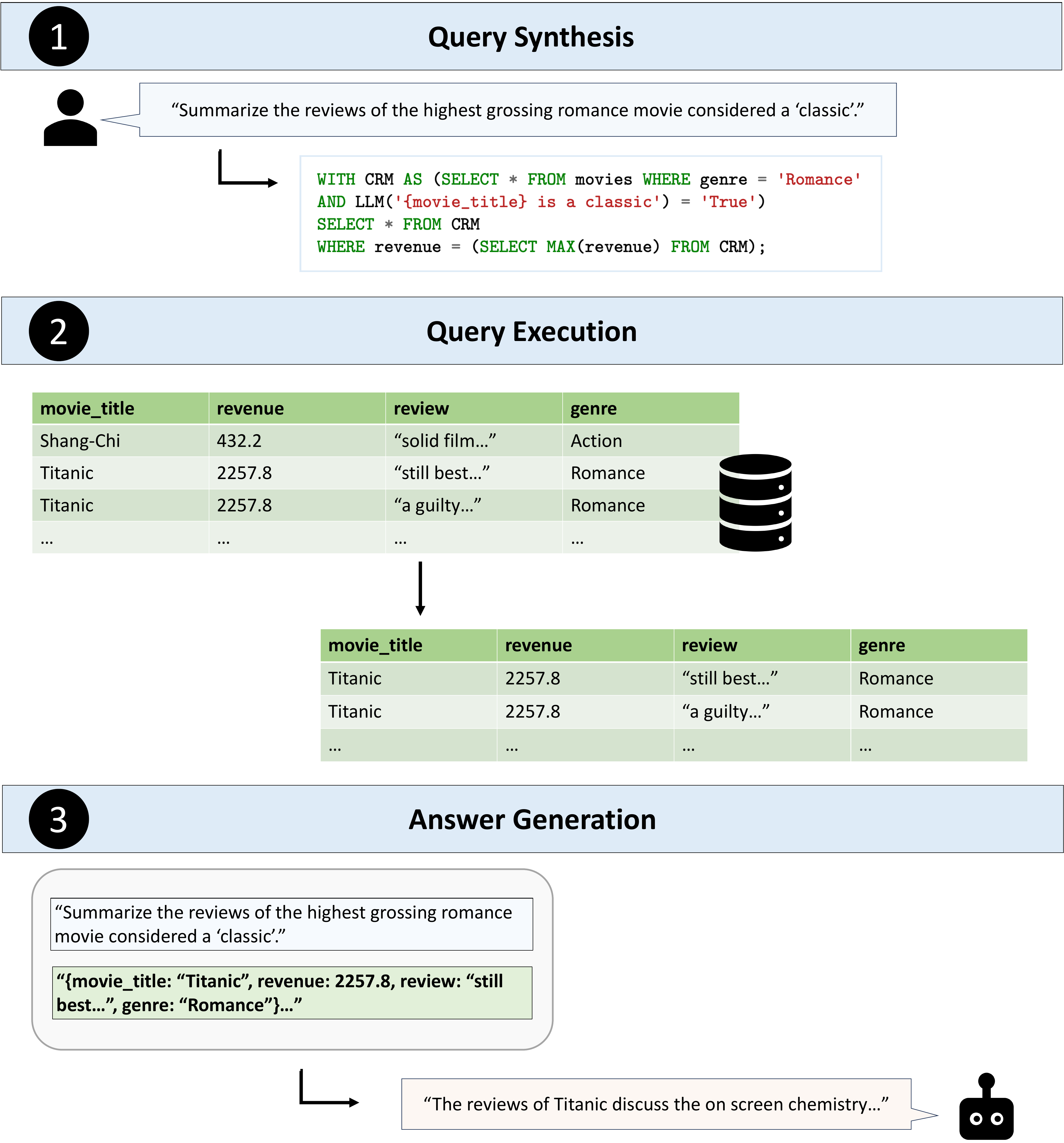}
  \caption{An example TAG implementation for answering the user's  natural language question over a table about movies. The TAG pipeline proceeds in three stages: query synthesis, query execution, and answer generation}
  \label{fig:example_tag}
   \vspace{-7mm}
\end{figure}

\subsection{Query Synthesis}
The \texttt{syn} function takes a natural language request $R$ and generates a query $Q$ to be executed by the database system. Given a user request, this step is responsible for \textbf{(a)} deducing which data is relevant to answering the request (e.g., using the table schema), and \textbf{(b)} performing semantic parsing to translate the user request into a query that can be executed by the database system. This query could be in any query language, but in our example we use SQL.

Figure \ref{fig:example_tag} shows an example TAG instantiation for the user query which asks, 
\emph{``Summarize the reviews of the highest grossing romance movie considered a ‘classic’"}. 
Here, the data source contains information about each movie's title, revenue, genre, and an associated review. In this step, the system leverages the semantic reasoning abilities of the LM to generate a SQL query that uses attributes on \verb|movie_title|, \verb|review|, \verb|revenue|, and \verb|genre| from the data source. Note that in this example, the database API is able to execute LM UDFs within SQL queries, so this step also introduces calls to the LM for each row to identify classic films within the query.

\subsection{Query Execution}
In the query execution step, the \texttt{exec} function executes the query $Q$ in the database system to obtain the table $T$.
This step leverages the database query engine to efficiently execute the query over vast amounts of stored data. The database API can be served by a wide variety of systems, which we explore in Section~\ref{sec:design_space}. Some APIs may allow for LM-based operators \cite{patel_lotus_2024, liu_suql_2024, liu_optimizing_2024, liu_declarative_2024}, permitting the database engine to leverage the LM's world knowledge and reasoning capabilities within \texttt{exec}.

In the example shown in Figure \ref{fig:example_tag}, the database query is a selection and ranking query written in SQL, which returns a table containing relevant rows. The query performs the selection using an LM to assess which movies are classics according to their \verb|movie_title|, as well as a standard filter on \verb|genre| to find romance movies. The query also ranks the results based on \verb|revenue| to find the highest grossing film. As the figure shows, the resulting table contains reviews for the movie ``Titanic''.

\subsection{Answer Generation}
The answer generation step in TAG mirrors the generation step in RAG.
In this step, the \texttt{gen} function uses the LM to generate an answer $A$ to the user's natural language request $R$, using the computed data $T$.

Figure~\ref{fig:example_tag} shows the final stage of the example TAG pipeline outputting a summary of the reviews on "Titanic" as the answer to the original user request. In the example, the relevant data $T$ is encoded as a string for the model to process. The encoded table is passed to the LM along with the original user request, $R$. To obtain the answer, this step leverages the model's semantic reasoning capabilities over the \verb|review| column to summarize the reviews.

%% file: Sections/Design-space.tex
\section{TAG Design Space}

\label{sec:design_space}
In this section, we explore the generality of the TAG model and describe the rich design space it produces, highlighting several under-studied opportunities for further research.

\heading{Query Types}
The TAG model is expressive enough to serve a broad range of natural language user queries. We consider two important query categorizations, according to \textbf{(a)} the \emph{level of data aggregation} required to answer the query and \textbf{(b)} the \emph{knowledge and capabilities} needed for answering the query. First, the TAG model captures both point queries, such as retrieval-based questions \cite{wu_stark_2024, zelikman_star_2022, lewis_retrieval-augmented_2021, gao_rarr_2023, anantha_open-domain_2021, lee_latent_2019} which require look-ups to one or a few rows of the database, as well as aggregation queries, such as summarization or ranking-based questions which require logical reasoning across many rows of the database. Secondly, the TAG model enables natural language queries with varying demands on the system to provide data or reasoning-based capabilities, including for tasks such as sentiment analysis and classification.

\heading{Data Model} The underlying data model can take many forms. 
Our implementation uses relational databases to store and retrieve structured attributes for knowledge-grounding in the downstream question-answering task. Others may operate on more unstructured data (e.g., free-text, images, video, and audio) or semi-structured data, which may be stored with a variety of data models, such as key-value, graph, vector, document, or object stores.

\heading{Database Execution Engine and API}
The underlying system used to store the data can use many possible database execution engines. Text2SQL considers the setting of an SQL-based query engine for retrieving relational data for user queries. In this setting, \texttt{syn} will leverage the knowledge of the data source, such as the table schema, and return a SQL query to perform the retrieval step. In another common setting, retrieval systems over vector embeddings, \texttt{syn} transforms the natural language query into an embedding and \texttt{exec} performs similarity-based retrieval over the vector store.

While these two settings have been widely studied, several under-studied alternative settings present interesting opportunities for efficiently implementing TAG systems to serve a broader range of queries. For instance, recent works augment relational models with semantic operators~\cite{patel_lotus_2024}, which provide a set of declarative AI-based operators (e.g., filtering, ranking, aggregating, and performing search with natural language specifiers) or LM user-defined functions~\cite{liu_optimizing_2024}, which provide a general-purpose LM function. Additionally, query languages like MADLib~\cite{hellerstein_madlib_2012}, Google's BigQuery ML~\cite{noauthor_introduction_nodate}, and Microsoft's Predictive SQL~\cite{williamdassafmsft_predict_2023} augment SQL-based APIs with native ML-based functions. Leveraging these systems provides unique opportunities for executing optimized reasoning-based retrieval patterns. For instance, in the example shown in Figure \ref{fig:example_tag}, a TAG pipeline implemented with semantic operators~\cite{patel_lotus_2024} might use a \verb|sem_filter| operator to filter rows based on whether they are a 'classic' during the query execution step.

\heading{LM Generation Patterns}
Given the table $T$ of relevant data, \texttt{gen} can be comprised from a vast array of implementation decisions to produce the final natural language answer $A$ in response to the user request $R$. While Text2SQL omits the final generation step and stops short after query execution, RAG pipelines typically leverage a single LM-call generation implementation where relevant data is fed in context. In this setting, several works study sub-problems related to table encoding~\cite{fang_large_2024}, prompt compression~\cite{chen_frugalgpt_2023}, and prompt tuning~\cite{khattab_dspy_2023} to optimize the in-context learning results.

More recent research, such as LOTUS~\cite{patel_lotus_2024}, highlights the potential of composing iterative or recursive LM generation patterns for answering queries involving reasoning-based transformations, aggregations, or rankings across multiple data rows. Early work demonstrates the rich design space presented by these LM-based algorithms and promising results on several downstream tasks.

%% file: Sections/Evaluation.tex
\section{Evaluation}
\label{sec:eval}

In this section, we introduce the first TAG benchmark and evaluate a collection of baselines, aiming to address the following questions:

\begin{enumerate}
    \item How do existing methods for table question answering perform on queries requiring \textit{semantic reasoning} or \textit{world knowledge}?
    \item How does a hand-written implementation of the TAG model, which divides computational and reasoning steps across DBMS and LM operations, perform on these queries?
\end{enumerate}

\subsection{Benchmark Methodology}
Existing benchmarks have explored how models perform on basic queries answerable entirely from data in the data source. We build upon prior work by modifying queries such that they require knowledge not directly available in the data source or semantic reasoning to answer. We select BIRD~\cite{bird}, a widely used Text2SQL benchmark on which LMs have been evaluated, for its large scale tables along with its variety of domains and query types.

\heading{Dataset} Our queries span 5 domains selected from BIRD, each containing diversity in query types. We select \textit{california\_schools}, \newline\textit{debit\_card\_specializing}, \textit{formula\_1}, \textit{codebase\_community}, and \textit{european\_football\_2} as the DB sources for our queries.

\heading{Queries} The BIRD benchmark defines fundamental query types, including match-based, comparison, ranking, and aggregation queries. We select queries among these types from the BIRD benchmark and modify them to require either world knowledge or semantic reasoning for the model to answer. As an example of a modified query requiring world knowledge, in the \textit{california\_schools} DB, a modified query adds an additional clause asking for only schools in the Bay Area. This information is not in the table and requires the model's world knowledge to answer. Next, a modified query requiring LM reasoning asks for the top 3 most sarcastic comments on a particular post in the \textit{codebase\_community} DB. For evaluation on these queries, we rely on human-labeled ground truth. Our final benchmark consists of 80 modified queries, 40 requiring parametric knowledge and 40 requiring reasoning, with 20 of each of the 4 chosen BIRD query types.

\heading{Evaluation metrics} We measure accuracy as the percentage of exact matches as compared to the labeled correct answer for the match-based, comparison, and ranking query types. For aggregation queries, we provide qualitative analysis on results using each baseline. We also measure execution time in seconds for each query.

\heading{Experimental setup} We use the instruction tuned variant of Meta's Llama-3.1 model~\cite{llama3} with 70B parameters as our LM for both Text2SQL and final output generation. We use SQLite3 as our database API for baselines involving SQL and use an E5 base embedding model~\cite{wang2024textembeddingsweaklysupervisedcontrastive} for our RAG baseline. We run Llama-3.1-70B-Instruct with vLLM~\cite{kwon2023efficientmemorymanagementlarge} on 8 A100 80GB GPUs.

\subsection{Baselines}
\heading{Text2SQL}
In this baseline, the LM generates SQL code which is run to obtain an answer. For a given NL query, we construct an LM prompt containing table schemas for every table in the query's domain, using the same prompt format as in the BIRD work. We evaluate this baseline executing the generated SQL code in SQLite3 and measuring the number of incorrect answers, including instances where the model fails to generate valid SQL code. 

\heading{Retrieval Augmented Generation (RAG)}
RAG style methods have been explored for table retrieval \cite{chen2024tableretrievalsolvedproblem, wu_stark_2024}, where tabular data is embedded into an index for search. For our baseline, we use \textit{row-level embeddings}. A given row is serialized as "- {col}: {val}" for each column before being embedded into a FAISS \cite{douze2024faiss} index. During query time, we perform vector similarity search to retrieve 10 relevant rows to feed in context to our model along with the NL question. 

\heading{Retrieval + LM Rank} We extend the RAG baseline by utilizing an LM to assign a score between 0 and 1 for retrieved rows to rerank rows before input to the model, as is done in the STaRK work \cite{wu_stark_2024}. We use Llama-3.1-70B-Instruct as our reranker.  

\heading{Text2SQL + LM}
In this baseline, our model is first asked to generate SQL to retrieve a set of relevant rows to answer a given NL query. This is an important distinction from the Text2SQL baseline, where the model is asked to directly generate SQL code that alone provides an answer to the query when executed. Similar to the RAG baseline, relevant rows are fed in context to the model once retrieved. 

\heading{Hand-written TAG}
We also evaluate hand-written TAG pipelines, which leverage expert knowledge of the table schema rather than automatic query synthesis from the natural language request to the database query. We implement our hand-written TAG pipelines with LOTUS \cite{patel_lotus_2024}. The LOTUS API allows programmers to declaratively specify query pipelines with standard relational operators as well as semantic operators, such as LM-based filtering, ranking, and aggregations. LOTUS also provides an optimized semantic query execution engine, which we use to implement the query execution and answer generation steps of our hand-written TAG pipelines. 

\subsection{Results}
\input{tables/results}
Table~\ref{tab:overall_match_comparison_ranking} shows the accuracy, measured by exact match, and execution time of each method. As the table shows, across the selected BIRD query types, we find that our hand-written TAG baseline consistently achieves 40\% exact match accuracy or better, where all other baselines fail to exceed 20\% accuracy.

The Text2SQL baseline performs poorly on all baselines with an execution accuracy no higher than 20\% but especially poorly on ranking queries with only 10\% accuracy, as many of the ranking queries require reasoning over text. The Text2SQL + LM generation baseline has similar poor performance across baselines, but does worse on match-based and comparison queries with only 10\% accuracy. On these query types, several context length errors occur trying to feed in many rows to the model after the executed SQL.

Turning our attention to the RAG baseline, we see that it fails to answer a single query correctly across all query types, highlighting its poor fit for queries in this space.
Adding LM reranking allows Retrieval + LM rank to answer a query correctly among the comparison queries, however the baseline still performs worse than all others besides RAG. 

Our hand-written TAG baseline answers 55\% of queries correctly overall, performing best on comparison queries with an exact match accuracy of 65\%. The baseline performs consistently well with over 50\% accuracy on all query types except ranking queries, due to the higher difficulty in ordering items exactly. Overall, this method gives us between a 20\% to 65\% accuracy improvement over the standard baselines.

Additionally, Table~\ref{tab:knowledge_reasoning} highlights the weaknesses of standard methods in answering the query types discussed in Section~\ref{sec:design_space}. Namely, vanilla Text2SQL especially struggles on queries requiring LM reasoning with 10\% exact match accuracy, due to its omission of the answer generation step. Meanwhile, the RAG baseline and Retrieval + LM Rank baseline struggle on all query types, answering only one query correctly, due to their reliance on the LM to handle all exact computation over data. In contrast, the hand-written TAG baseline achieves over 50\% accuracy on both queries requiring knowledge and queries requiring reasoning, emphasizing the TAG model's versatility in the queries it encapsulates. 

Notably, along with offering superior accuracy, the hand-written TAG method offers an efficient implementation with up to 3.1$\times$ lower execution time over other baselines. The hand-written baseline takes an average of 2.94 seconds for all queries. This relatively low execution time highlights that an efficient TAG system can be designed by exploiting efficient batched inference of LMs. 

Lastly, we qualitatively analyze the results of each baseline on aggregation queries. Figure~\ref{fig:agg_results} shows the results for the RAG, Naive TAG, and hand-written baselines on the example query \textit{"Provide information about the races held on Sepang International Circuit."}. The RAG baseline is only able to provide information about some of the races, as most of the relevant races are not retrieved. On the other hand, the Text2SQL + LM baseline is not able to utilize any information from the DBMS, relying only on parametric knowledge and providing no further analysis. The hand-written baseline provides a thorough summary of all the races from 1999 to 2017 held at Sepang International Circuit. We observe a similar trend across other aggregation queries provided by the benchmark, with initial results highlighting the potential of TAG systems to successfully aggregate large amounts of data to provide informative answers. We leave quantitative analysis to future work.

%% file: tables/results.tex
\setlength{\tabcolsep}{3.75pt}
\begin{table*}[h]
\begin{threeparttable}
\centering
\caption{Accuracy and execution time (ET) for TAG benchmark queries, averaged over all queries and each query type: TAG significantly improves answer quality while achieving the fastest or nearly fastest execution time.
}
\vspace{-.3cm}
\small
\begin{tabular}{l | cc | cc | cc | cc | cc}
\toprule    
 \multirow{2}{*}{Method} & \multicolumn{2}{c}{Overall} & \multicolumn{2}{c}{Match-based} & \multicolumn{2}{c}{Comparison} & \multicolumn{2}{c}{Ranking} & \multicolumn{2}{c}{Aggregation} \\
\cmidrule(lr){2-3} \cmidrule(lr){4-5}\cmidrule(lr){6-7} \cmidrule(lr){8-9} \cmidrule(lr){10-11}
& Exact Match $\uparrow$\tnote{1}  & ET (s) $\downarrow$ & Exact Match $\uparrow$  & ET (s) $\downarrow$ & Exact Match $\uparrow$  & ET (s) $\downarrow$ & Exact Match $\uparrow$  & ET (s) & Exact Match $\uparrow$  & ET (s) $\downarrow$ \\
\midrule
Text2SQL & 0.17 & 5.63 & 0.20 & 4.72 & 0.20 & 4.01 & 0.10 & 7.26 & N/A & 6.53 \\
RAG & 0.00 & 3.23 & 0.00 & 3.73 & 0.00 & \textbf{2.29} & 0.00 & \textbf{2.01} & N/A & 4.89  \\
Retrieval + LM Rank & 0.02 & 4.82 & 0.00 & 6.20 & 0.05 & 4.19 & 0.00 & 3.42 & N/A & 5.46 \\
Text2SQL + LM & 0.13 & 9.08 & 0.10 & 11.25 & 0.10 & 3.89 & 0.20 & 11.80 & N/A & 9.38 \\
Hand-written TAG & \textbf{0.55} & \textbf{2.94} & \textbf{0.60} & \textbf{1.70} & \textbf{0.65} & 5.05 & \textbf{0.40} & 2.50 & N/A & \textbf{2.50} \\
\bottomrule
\end{tabular}
\label{tab:overall_match_comparison_ranking}
\begin{tablenotes}
\item[1] Excludes aggregation since accuracy is not measured.
\end{tablenotes}
\end{threeparttable}
\end{table*}

\setlength{\tabcolsep}{3pt}
\begin{table}[h]
\centering
\caption{TAG benchmark results averaged over queries requiring Knowledge or Reasoning: TAG performs consistently well with above 50\% exact match accuracy on both Knowledge and Reasoning query types.
}
\vspace{-.3cm}
\small
\begin{tabular}{l| cc | cc}
\toprule    
\multirow{2}{*}{Method} & \multicolumn{2}{c}{Knowledge} & \multicolumn{2}{c}{Reasoning} \\
\cmidrule(lr){2-3} \cmidrule(lr){4-5}
& Exact Match $\uparrow$  & ET (s) $\downarrow$ & Exact Match $\uparrow$  & ET (s) $\downarrow$ \\
\midrule
Text2SQL & 0.20 & 5.23 & 0.10 & 5.52 \\
RAG & 0.00 & \textbf{2.73} & 0.00 & 2.58 \\
Retrieval + LM Rank & 0.03 & 4.97 & 0.00 & 3.87 \\
Text2SQL + LM & 0.10 & 10.27 & 0.20 & 6.39 \\
Hand-written TAG & \textbf{0.53} & 3.50 & \textbf{0.60} & \textbf{2.24} \\
\bottomrule
\end{tabular}
\label{tab:knowledge_reasoning}
\end{table}

\begin{figure*}[h]
  \centering
  \includegraphics[width=\textwidth]{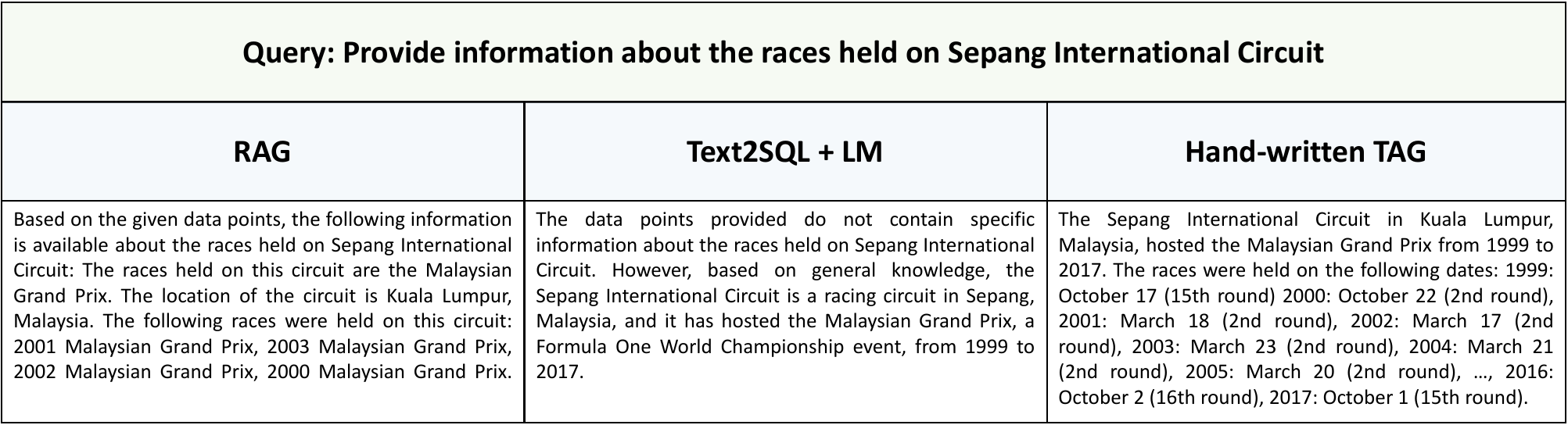}
  \caption{Example Aggregation Results: The RAG baseline provides an incomplete answer to the query while Text2SQL + LM fails to answer the question using any data from the DB. The Hand-written TAG baseline provides the most thorough answer, synthesizing data from the DB and its own world knowledge.}
  \label{fig:agg_results}
\end{figure*}

%% file: Sections/RelatedWork.tex
\section{Related Work}
\label{sec:related_work}
\heading{Text2SQL} Text2SQL using LMs has been extensively explored in prior work. WikiSQL \cite{wikisql}, Spider \cite{spider}, and BIRD \cite{bird} are all popular datasets for cross-domain Text2SQL. These datasets contain structured data across many domains on which the task of converting natural language queries to SQL is evaluated. However, this direction does not utilize model capabilities beyond SQL generation, keeping queries that require reasoning or knowledge beyond a static data source out of scope. 

\heading{Retrieval Augmented Generation} Retrieval augmented generation (RAG) \cite{lewis_retrieval-augmented_2021} enables LMs to extend beyond their parametric knowledge to large collections of text. SQuAD \cite{squad} and HotPotQA \cite{hotpot} focus on question-answering over single document and multiple document sources respectively. The dense table retrieval (DTR) model \cite{DTR} extends RAG to tabular data, embedding tabular context to retrieve relevant cells and rows for a query. Join-aware table retrieval \cite{chen2024tableretrievalsolvedproblem} adds a table-table similarity score term to the DTR model to improve performance on complex queries involving joined tables. In contrast to prior RAG work, the TAG model encompasses a larger field of queries users have over their data by leveraging LM capabilities in the query execution step and allowing DBMS operations for exact computation over large amounts of data. 

\heading{NL Queries over Semi-structured Data} Prior work has explored the relational information between table entities and unstructured entity fields in semi-structured data sources. STaRK \cite{wu_stark_2024} evaluates table retrieval methodologies across semi-structured knowledge bases (SKBs), including both structural and nonstructural information. SUQL \cite{liu_suql_2024} addresses the task of conversational search, where an LM is used as a semantic parser to handle unstructured components user queries over hybrid data. While these works primarily focus on natural language \textit{search} queries over semi-structured data, we seek to explore a broader range of queries leveraging more LM capabilities for tasks beyond search and lookup. 

\heading{Agentic Data Assistants} Recent work has explored LM agents as data assistants \cite{hu2024infiagentdabenchevaluatingagentsdata}. Spider2-V \cite{cao2024spider2vfarmultimodalagents} explores multimodal agent performance in tasks involving code generation and GUI controls. Though we define the TAG model tractably as one iteration of the \texttt{syn}, \texttt{exec}, and \texttt{gen} functions, future work may explore extending this in an agentic loop.

%% file: Sections/Conclusion.tex
\section{Conclusion}
In this work we proposed table-augmented generation (TAG) as a unified model for answering natural language questions over databases. We developed benchmarks to study two important types of queries: those that require world knowledge, and those that require semantic reasoning capabilities. Our systematic evaluation confirms that baseline methods are unable to make meaningful traction on these tasks. However, hand-written TAG pipelines can achieve up to $65\%$ higher accuracy, demonstrating substantial research opportunities for building TAG systems.

%% file: Sections/Acknowledgements.tex
\section{Acknowledgements}
This research was supported in past by affiliate members and supporters  of the Stanford DAWN project and the Sky Computing Lab at Berkeley, including Accenture, AMD, Anyscale, Cisco, Google, IBM, Intel, Meta, Microsoft, Mohamed Bin Zayed University of Artificial Intelligence, NVIDIA, Samsung SDS, SAP, VMware, and a Sloan Fellowship. Any opinions, findings, and conclusions or recommendations expressed in this material are those of the authors and do not necessarily reflect the views of the sponsors.

%% file: Sections/Appendix.tex
\section{Sample Queries}
We detail the modifications made to BIRD queries for our benchmark. Each query is modified to require either LM knowledge or reasoning to answer. Sample queries are shown below.

\input{tables/queries}

\section{LM Prompts}
We summarize the prompts used with instruction tuned Llama-3.1 80B for query synthesis and answer generation. 

\subsection{Query Synthesis}
For the query synthesis step, in our case a Text2SQL step, we use the same table schema encoding and LM prompt as the original BIRD benchmark. An example prompt for query synthesis is shown below. 

\begin{lstlisting}
-- Example Prompt for Query Synthesis
CREATE TABLE frpm
(
    CDSCode TEXT not null primary key,
    Academic Year TEXT null,
    ...
)

CREATE TABLE satscores
(
    ...
    AvgScrRead INTEGER null,
    AvgScrMath INTEGER null,
    ...
)

CREATE TABLE schools
(
    ...
    District TEXT not null,
    School TEXT null,
    ...
)

-- External Knowledge: None
-- Using valid SQLite and understading External Knowledge, answer the following questions for the tables provided above.
-- Among the schools with the average score in Math over 560 in the SAT test, how many schools are in the bay area?
SELECT

\end{lstlisting}

\subsection{Answer Generation}
On the Text2SQL + LM and RAG baselines, the answer generation step requires the LM to answer a user question with the provided rows in context. We utilize a separate prompt for aggregation queries, while match-based, comparison, and ranking share the same prompt. We show both prompts below.

\begin{lstlisting}
-- Example Prompt for Answer Generation for Match-based, Comparison, and Ranking Queries.

You will be given a list of data points and a question. Use the data points to answer the question. Your answer must be a list of values that is evaluatable in Python. Respond in the format [value1, value2, ..., valueN]. If you are unable to answer the question, respond with []. Respond with only the list of values and nothing else. If a value is a string, it must be enclosed in double quotes.

Data Point 1:
...
- School: <school name for first row>
- AvgScrMath: <average math score first row>
...
Data Point 2:
...
- School: <school name for second row>
- AvgScrMath: <average math score second row>
...

Question: Among the schools with the average score in Math over 560 in the SAT test, how many schools are in the bay area?
\end{lstlisting}

\begin{lstlisting}
-- Example Prompt for Answer Generation for Aggregation Queries.

You will be given a list of data points and a question. Use the data points to answer the question. If a value is a string, it must be enclosed in double quotes.

Data Point 1:
...
- PostId: <post id for first row>
- Text: <comment text first row>
...
Data Point 2:
...
- PostId: <post id for second row>
- Text: <comment text second row>
...

Question: Summarize the comments made on the post titled "How does gentle boosting differ from AdaBoost?" to answer the original question.
\end{lstlisting}

\section{Handwritten Pipelines}
We use the LOTUS package to construct hand-written TAG pipelines. For each query in our benchmark, a pipeline consisting of a series of dataframe transformations and filters along with LOTUS semantic LM operators was written in Python. Example pipelines are visible below.

\begin{lstlisting}
-- Match-based
query = "What is the grade span offered in the school with the highest longitude in cities in that are part of the 'Silicon Valley' region?"

schools_df = pd.read_csv("../pandas_dfs/california_schools/schools.csv")
unique_cities = pd.DataFrame(schools_df["City"].unique(), columns=["City"])
sv_cities = unique_cities.sem_filter("{City} is a city in the Silicon Valley region")
schools_df = schools_df[schools_df["City"].isin(sv_cities["City"])]
schools_df = schools_df.sort_values(by=["Longitude"], key=abs, ascending=False).head(1)
prediction = schools_df["GSoffered"].tolist()[0]

return prediction
\end{lstlisting}
\begin{lstlisting}
-- Ranking
query = "Of the 5 posts wih highest popularity, list their titles in order of most technical to least technical."

posts_df = (
    pd.read_csv("../pandas_dfs/codebase_community/posts.csv").sort_values(by=["ViewCount"], ascending=False).head(5)
)

prediction = posts_df.sem_topk("What {Title} is most technical?", 5).Title.values.tolist()

return prediction
\end{lstlisting}
\begin{lstlisting}
-- Aggregation
query = "Summarize the comments made on the post titled 'How does gentle boosting differ from AdaBoost?' to answer the original question"

posts_df = pd.read_csv("../pandas_dfs/codebase_community/posts.csv")
comments_df = pd.read_csv("../pandas_dfs/codebase_community/comments.csv")
posts_df = posts_df[posts_df["Title"] == "How does gentle boosting differ from AdaBoost?"]
merged_df = pd.merge(posts_df, comments_df, left_on="Id", right_on="PostId")

prediction = merged_df.sem_agg("Summarize the comments", all_cols=True)._output[0]
return prediction
\end{lstlisting}

%% file: tables/queries.tex
\begin{table}[htbp]
    \centering
    \begin{tabularx}{\textwidth}{X}
        \toprule
        \multicolumn{1}{c}{\textbf{Match-based}} \\
        \midrule
        \textbf{Original BIRD Query:} \\
        \textit{What is the grade span offered in the school with the highest longitude?} \\
        \textbf{Modified Query:} \\
        \textit{What is the grade span offered in the school with the highest longitude \textcolor{orange}{in cities in that are part of the 'Silicon Valley' region}?} \\
        \textbf{Analysis:} \\
        This query is modified to require LM knowledge of which cities are within the Silicon Valley region of California, information not available in the data source. \\
        \midrule
        \multicolumn{1}{c}{\textbf{Comparison}} \\
        \midrule
        \textbf{Original BIRD Query:} \\
        \textit{Among the players whose height is over 180, how many of them have a volley score of over 70?} \\
        \textbf{Modified Query:} \\
        \textit{Among the players whose height is over 180, how many of them have a volley score of over 70 \textcolor{orange}{and are taller than Stephen Curry}?} \\
        \textbf{Analysis:} \\
        This query is modified to require LM knowledge of how tall Stephen Curry is. \\
        \midrule
        \multicolumn{1}{c}{\textbf{Ranking}} \\
        \midrule
        \textbf{Original BIRD Query:} \\
        \textit{What are the titles of the top 5 posts with the highest popularity?} \\
        \textbf{Modified Query:} \\
        \textit{Of the 5 posts wih highest popularity, \textcolor{orange}{list their titles in order of most technical to least technical}.} \\
        \textbf{Analysis:} \\
        This query is modified to require LM reasoning over a textual field, the post's title. \\
        \midrule
        \multicolumn{1}{c}{\textbf{Aggregation}} \\
        \midrule
        \textbf{Original BIRD Query:} \\
        \textit{Write all comments made on the post titled 'How does gentle boosting differ from AdaBoost?'} \\
        \textbf{Modified Query:} \\
        \textit{\textcolor{orange}{Summarize} the comments made on the post titled 'How does gentle boosting differ from AdaBoost?' to answer the original question.} \\
        \textbf{Analysis:} \\
        This query is modified to rely on LM reasoning over text on the textual comment fields to provide a summary. \\
        \bottomrule
    \end{tabularx}
\end{table}